\documentclass[aps,prl,reprint,groupedaddress]{revtex4-1}
\usepackage{bm}
\usepackage{graphicx}

\begin{document}


\title{Third Stable Branch and Tristability\\
of Nuclear Spin Polarization in Single Quantum Dot System}


\author{S.~Yamamoto}
	\affiliation{Division of Applied Physics, Hokkaido University, N13 W8, Kitaku, Sapporo 060-8628, Japan}
\author{R.~Kaji}
	\affiliation{Division of Applied Physics, Hokkaido University, N13 W8, Kitaku, Sapporo 060-8628, Japan}
\author{H.~Sasakura}
	\affiliation{Division of Applied Physics, Hokkaido University, N13 W8, Kitaku, Sapporo 060-8628, Japan}
\author{S.~Adachi}
\email{adachi-s@eng.hokudai.ac.jp}
	\affiliation{Division of Applied Physics, Hokkaido University, N13 W8, Kitaku, Sapporo 060-8628, Japan}


\date{\today}

\begin{abstract}
Semiconductor quantum dots provide a spin-coupled system of an electron and nuclei via enhanced hyperfine interaction.
We showed that the nuclear spin polarization in single quantum dots can have three stable branches under a longitudinal magnetic field.
The states were accompanied by hysteresis loops around the boundaries of each branch with a change in the excitation condition.
To explain these findings, we incorporated the electron spin relaxation caused by the nuclear spin fluctuation into the previously-studied dynamic nuclear spin polarization mechanism.
The model reproduces the new features of nuclear spin polarization and the associated strong reduction in the observed electron spin polarization, and can refer to the tristability of nuclear spin polarization.
\end{abstract}


\maketitle


In semiconductor quantum dots (QDs), the strong localization of electron wave function enhances hyperfine interaction (HFI).
Simultaneous spin-flip via HFI transfers the spin angular momentum from an electron to the lattice nuclear spin ensemble, and thus, nuclear spin polarization (NSP) can be established by spin-polarized electron injection~\cite{Gammon01, Yokoi05, Eble06}.
The unparalleled spin coherence of nuclei has always been attractive for application to quantum memory~\cite{Taylor03}, and recent demonstrations and predictions using spin waves suggest the possibility of quantum read-write processes with over 90 \% accuracy~\cite{Gangloff19,Denning19}.
On the other hand, the nuclear spin fluctuation causes a serious electron spin decoherence, which is an undesired aspect of spin coupling via HFI~\cite{Merkulov02}. 
Recent studies have reported that this electron spin decoherence can be eliminated by using the spin refocusing technique~\cite{Botzem16,Stockill16}.

One of the most interesting properties of NSP is bistability; the NSP transits abruptly between two stable coexisting branches due to the negative and positive feedbacks of the spin transfer rate~\cite{Braun06, Tartakovskii07, Maletinsky07, Kaji08, Belhadj08}.
This phenomenon occurs when electron spin splitting, which limits the spin transfer rate, is reduced by the compensation of the external magnetic field by an effective field originating from NSP (i.e. nuclear field).

In this study, we show that NSP in individual QDs potentially has three stable branches.
This intriguing behavior was demonstrated by steady-state photoluminescence (PL) measurements under non-resonant excitation, and it can be explained by a phenomenological rate equation, including the effects of nuclear spin fluctuation.
In addition, the proposed model predicts that a QD spin system exhibits a \textit{tristable} response under some proper conditions.
Our findings remind us of the degree of complication of spin coupling via HFI in a QD, which sometimes causes unintuitive behaviors, such as the bidirectional NSP formation~\cite{Latta09,Hogele12} and the anomalous Hanle effect~\cite{Krebs10,Yamamoto18}.
New knowledge related to nuclear spin fluctuation contributes to a deeper understanding of electron spin decoherence and a hybrid quantum system composed of an electron and nuclei.


We used single In$_{0.75}$Al$_{0.25}$As/Al$_{0.3}$Ga$_{0.7}$As self-assembled (SA) QDs grown by molecular-beam epitaxy on (100) GaAs substrate. 
We analyzed the PL spectra of three different single QDs (QD1, QD2, and QD3) at 6 K. 
To explore the nonlinearity of NSP in the QDs, a static magnetic field $B_z$ up to 5 T was applied along the sample growth direction ($z$). 
A circularly polarized excitation was employed to inject the spin-polarized electron and hole into the wetting layer of the QDs (${\sim} 1.6986$ eV).
Here, we define the degree of circular polarization as $\rho_{\rm c} {=} (I^- {-} I^+) / (I^- {+} I^+)$, where $I^{-(+)}$ is the intensity of $\sigma^{-(+)}$ polarized PL.
In this work, we focus on a positively charged exciton $X^+$ that consists of one electron and two holes in a spin singlet.
Since $\rho_{\rm c}$ depends only on the electron spin projection on $z$-axis, $\langle{S_z}\rangle$, in the case of $X^+$, $\rho_{\rm c}$ is related directly with $\langle S_z \rangle$ as $\rho_{\rm c} {=} 2\langle S_z \rangle$.
The established NSP $\langle I_z \rangle$ along the $z$-axis was monitored through the Overhauser shift $\Delta_{\rm OS}$ defined as $2\tilde{A}\langle {I_z} \rangle$, where $\tilde{A}$ is a hyperfine constant ${\sim} 50\ {\rm \mu eV}$~\cite{Testelin09, Chekhovich17}. 
The details of the sample and measurements are presented in Ref.~\cite{SupplMat}.

\begin{figure}
\includegraphics[width=\hsize]{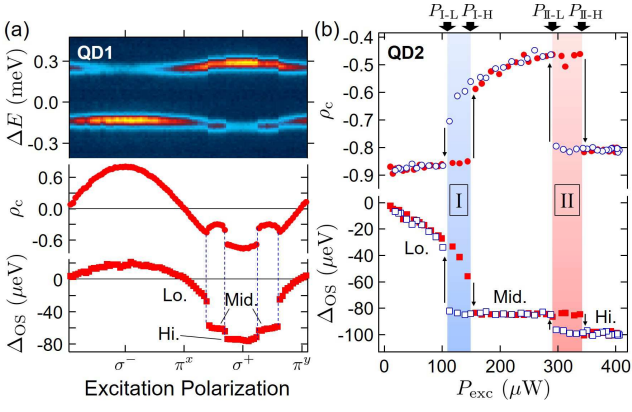}
\caption{(a) Excitation polarization dependence of $X^+$ PL from QD1 at $B_z{=}{+}3.0$ T and 6 K. Top panel is a color-scale plot of PL spectra whose vertical axis $\Delta E$ indicates the PL energy from 1.6264 eV. $\rho_{\rm c}$ (middle) and $\Delta_{\rm OS}$ (bottom) are obtained from the data in the top panel by the spectral fitting. (b) $P_{\rm exc}$ dependences of $\rho_{\rm c}$ (top) and $\Delta_{\rm OS}$ (bottom) of QD2 at $B_z{=}{+}5.0$ T under $\sigma^+$ excitation. Filled (open) symbols indicate the results with increasing (decreasing) $P_{\rm exc}$. \label{Fig_Experiment1}}
\end{figure}

The excitation polarization dependence of the $X^+$ PL spectra obtained from QD1 at $B_z{=}{+}3.0$ T is shown in the top panel of Fig.~\ref{Fig_Experiment1}(a) as a color-scale plot.
Because the g factor ${\rm g}_{\rm e}$ of the conduction electron is positive in our InAlAs QDs~\cite{Kaji19}, the compensation of $B_z$ by a nuclear field occurred under $\sigma^+$ excitation, where the spin-down electron is photo-injected selectively.
In case of such a condition, the PL energies and intensities indicated abrupt changes.
To closely examine this point, both the $\rho_{\rm c}$ and $\Delta_{\rm OS}$ that are yielded from the spectra are plotted in the middle and bottom panels, respectively.
It should be noted that there are two-stage abrupt jumps, which suggest three stable branches labeled Lo., Mid., and Hi. in the figure.

We also found three of such stable branches of NSP by changing the excitation power $P_{\rm exc}$.
Figure~\ref{Fig_Experiment1}(b) shows the $\rho_{\rm c}$ and $\Delta_{\rm OS}$ obtained from QD2 at $B_z{=}{+}5.0$ T under $\sigma^+$ excitation.
In the figure, the filled and open circles (squares) indicate the observed $\rho_{\rm c}$ ($\Delta_{\rm OS}$) with increasing and decreasing $P_{\rm exc}$, respectively.
With increasing $P_{\rm exc}$, $|\rho_{\rm c}|$ and $|\Delta_{\rm OS}|$ jumped simultaneously at approximately 150 $\mu$W ($P_{\rm I\mathchar`-H}$) and an additional abrupt change occurred at approximately 350 $\mu$W ($P_{\rm I\!I\mathchar`-H}$) again.
The similar behavior with two critical points ($P_{\rm I\!I\mathchar`-L}$ and $P_{\rm I\mathchar`-L}$) was observed with decreasing $P_{\rm exc}$.
Notably, $P_{\rm I(I\!I) \mathchar`-L}$ was lower than $P_{\rm I(I\!I) \mathchar`-H}$.
That is to say,  hysteresis loops were observed around each boundary of the branches [loop I (II) for the boundary between the Lo. and Mid. (the Mid. and Hi.) -branches].

\begin{figure}
\includegraphics[width=\hsize]{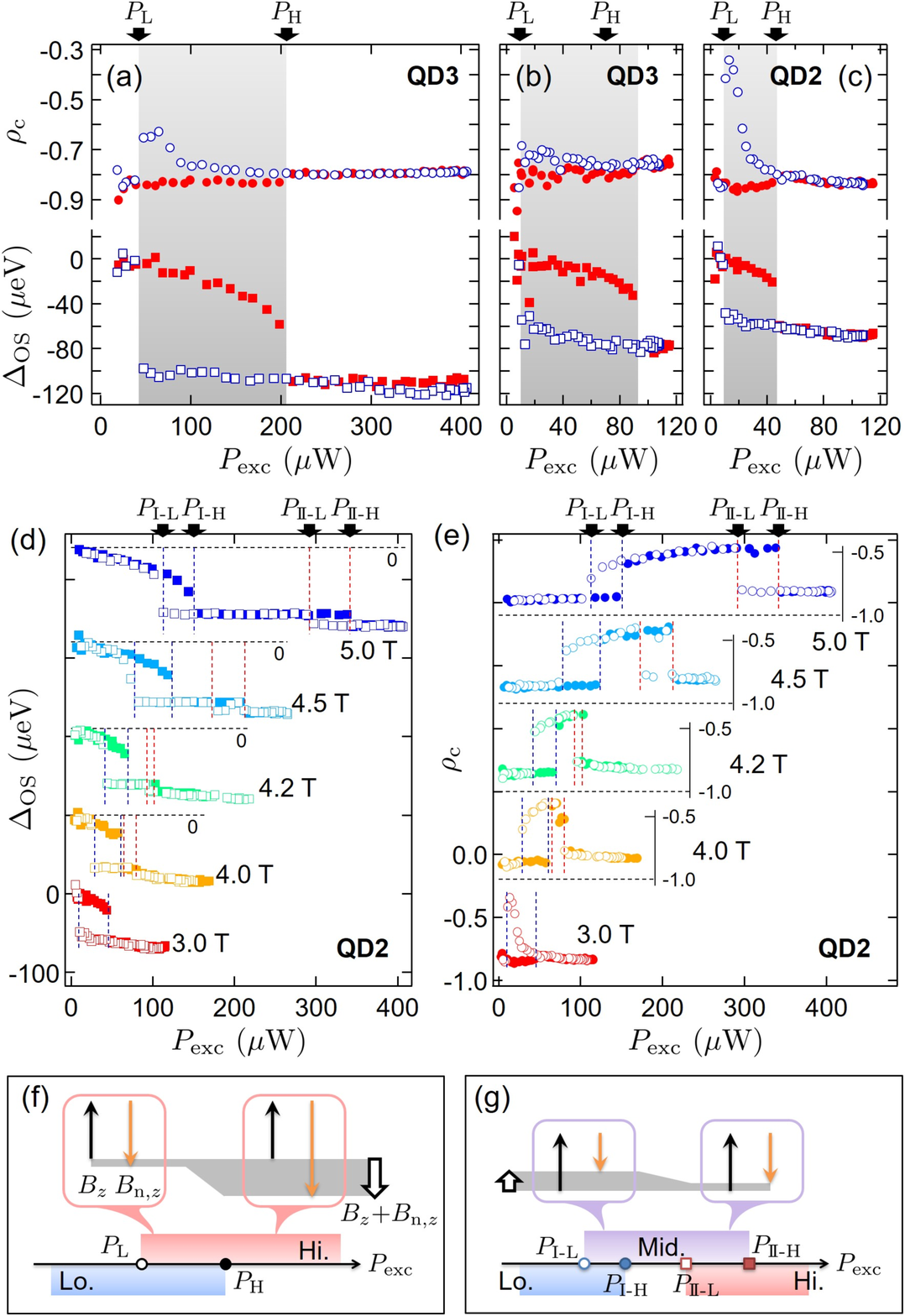}
\caption{(a)--(c) $P_{\rm exc}$ dependences of $\rho_{\rm c}$ and $\Delta_{\rm OS}$ (top and bottom panels) under $\sigma^+$ excitation. (a) and (b) are observed in QD3 at $B_z{=}{+}5.0$ T and +3.0 T, respectively, and (c) is observed in QD2 at $B_z{=}{+}3.0$ T. (d),  (e) $P_{\rm exc}$ dependences of $\Delta_{\rm OS}$ and $\rho_{\rm c}$ of QD2 at various $B_z$. The data are shifted vertically for clarity. (f), (g) Schematics of effective fields on an electron in the Hi.-branch of the bistable case (f) and in the Mid.-branch of three stable branches case (g). Blue, purple, and red shades indicate the $P_{\rm exc}$ ranges of the Lo., Mid., and Hi.-branches, respectively. Thin black and orange arrows depict $B_z$ and $B_{{\rm n},z}$, respectively. 
The sum, $B_{z}{+}B_{{\rm n}, z}$ is represented by the open arrows, and the gray shades indicate the magnitude and sign of the arrows.
\label{Fig_Experiment2}}
\end{figure}

It should be noted that the Mid.-branch is not always observed; its appearance depends on the QD properties and experimental conditions.
For example, Fig.~\ref{Fig_Experiment2}(a) shows $P_{\rm exc}$ dependences of the $\rho_{\rm c}$ and $\Delta_{\rm OS}$ of QD3 at $B_z{=}{+}5.0$ T.
Although the experimental condition was the same as in the case of Fig.~\ref{Fig_Experiment1}(b)~\cite{SupplMat}, the Mid.-branch did not appear, and only a single hysteresis loop was observed. 
Further, QD2 revealed the bistability in $B_z{=}{+}3.0$ T, as shown in Fig.~\ref{Fig_Experiment2}(c), as well as QD3 in Fig.~\ref{Fig_Experiment2}(b).
Figures~\ref{Fig_Experiment2}(d) and (e) show the $P_{\rm exc}$ dependences of $\Delta_{\rm OS}$ and $\rho_{\rm c}$ at various $B_z$ obtained from QD2.
The width of loop II and the region corresponding to the Mid.-branch reduced with decreasing $B_{z}$, and the Mid.-branch disappeared completely at a $B_{z}$ smaller than +4.0 T.
Although the presence of the Mid.-branch could not be judged distinctively from the change in $\Delta_{\rm OS}$ at approximately +4.0 T, a reduction of $|\rho_{\rm c}|$ in the Mid.-branch was larger compared with the other two and was the most prominent indication of the Mid.-branch.

To examine the difference between the bistable and the three stable cases, we focus on the compensation point and the magnitude relation between the nuclear field $B_{{\rm n},z}$ and $B_z$ in each branch.
The compensation point, where $B_{{\rm n},z}{+}B_z$ becomes zero, is explicitly reflected on the reduction of $|\rho_{\rm c}|$; hence, $|\langle S_z \rangle|$.
This is because, the electron spin relaxation, which is strongly suppressed by the Zeeman splitting, is enhanced at the compensation condition due to the degeneracy of electron Zeeman states.

In the bistable case, as shown in Fig.~\ref{Fig_Experiment2}(a)--(c), the lowest value of $|\rho_{\rm c}|$ was realized at $P_{\rm L}$, where $|\Delta_{\rm OS}|$ drops with decreasing $P_{\rm exc}$.
Thus, the $P_{\rm L}$ of the Hi.-branch was considered to be the compensation point. 
Since $B_{{\rm n},z}$ usually continues to grow with increasing $P_{\rm exc}$, as long as the system stays the same branch, the relation $|B_{{\rm n}, z}| {>} |B_{z}|$ is held in the region $P_{\rm exc} {>} P_{\rm L}$.
Accordingly, the increment of $P_{\rm exc}$ results in the increase of the total field seen by a QD electron [shown as a gray shade in Fig.~\ref{Fig_Experiment2}(f)], and the restoration of $\langle S_z \rangle$ in the Hi.-branch with increasing $P_{\rm exc}$ is expected.
This can be observed in Fig.~\ref{Fig_Experiment2}(a)--(c).

In the case with three stable branches, as shown in Fig.~\ref{Fig_Experiment1}(b), on the other hand, the lowest value of $|\rho_{\rm c}|$, and hence the compensation point, appeared at $P_{\rm I\!I\mathchar`-H}$.
At this point, $\Delta_{\rm OS}$ changes abruptly from the Mid. to the Hi.-branch with increasing $P_{\rm exc}$.
Throughout the Mid.-branch, since $|\rho_{\rm c}|$ decreased with increasing $P_{\rm exc}$, the electron Zeeman splitting was considered to keep reducing as $P_{\rm exc}$ approached $P_{\rm I\!I\mathchar`-H}$.
This situation is achieved if $|B_{{\rm n},z}| {<} |B_z|$ in the Mid.-branch, as illustrated in Fig.~\ref{Fig_Experiment2}(g).
Therefore, the magnitude relation between $|B_{{\rm n},z}|$ and $|B_z|$ in the Mid.-branch was opposite to that in the Hi.-branch.

Summarizing the observed data, the emergence of the Mid.-branch seemed to depend on the properties of the QD system. 
Comparing Figs.~\ref{Fig_Experiment2}(b) and (c), for example, although both data showed a single hysteresis loop under the same $B_{z}$, the widths were significantly different, which could be attributed to the difference in the correlation time of the HFI and/or relaxation time of the NSP.
In addition, the reduction of $|\rho_{\rm c}|$ at the compensation point was more significant in QD2 compared with that in QD3. 
This indicates that the electron spin relaxation rate in QD2 under the condition of $B_z{+}B_{{\rm n}, z} {=}0$ is larger than that in QD3.

\begin{figure}
\includegraphics[width=\hsize]{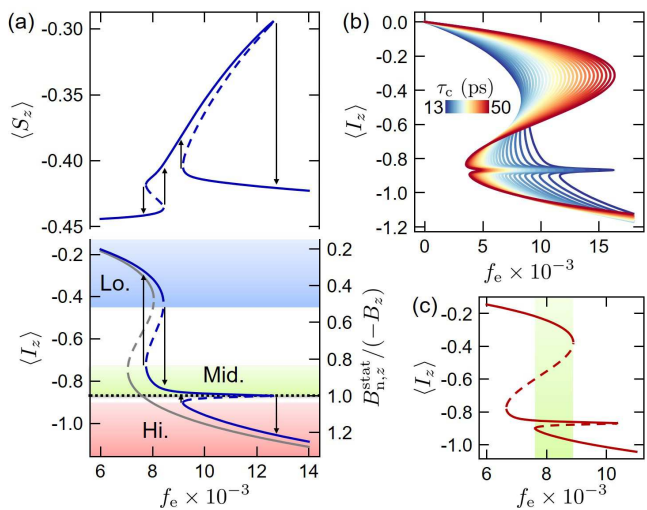}
\caption{
(a) Calculated $\langle S_z \rangle$ (top) and $\langle I_z \rangle$ (bottom) at $B_z{=+}5.0$ T. The used parameters are presented in Ref.~\cite{SupplMat}. 
The horizontal axis $f_{\rm e}$ is the QD occupation factor, which corresponds to $P_{\rm exc}$ in the experiment. 
The dashed part indicates an unstable solution, which is hard to access experimentally.
In the bottom panel, the gray curve indicates the calculation with the conventional model. 
The right vertical axis indicates corresponding $B_{{\rm n},z}^{\rm stat}$ normalized by ${-}B_z$, where the compensation condition (the dotted line) is unity. 
(b) Calculated $\langle I_z \rangle$ with various $\tau_{\rm c}$
(c) Calculated $\langle I_z \rangle$ with moderate $\tau_{\rm c}$. 
Tristability appears in the highlighted region.
\label{Fig_Calc}}
\end{figure}

To explain the observed Mid.-branch and the aforementioned dependences on QD properties, we introduce an effect of nuclear spin fluctuation known as \textit{frozen fluctuation}~\cite{Merkulov02} in addition to the conventional treatment of NSP~\cite{Eble06, Abragam, Krebs08, Urbaszek13}.
In accordance with  Kuznetsova \textit{et al.}~\cite{Kuznetsova13}, the nuclear field is considered as the sum of a static part $\bm{B}_{\rm n}^{\rm stat}$ and a fluctuation part $\bm{B}_{\rm f}$.
From the viewpoint of energy conservation of the flip-flop process via HFI, the formation rate of NSP is enhanced around the compensation condition $B_{z}{+}B_{{\rm n},z}^{\rm stat}{=}0$.
However, nuclear spin fluctuation causes the deterioration of the averaged electron spin polarization $\langle S_z \rangle$ around the compensation condition~\cite{Merkulov02}, and thus reduces the NSP formation rate.
Assuming an isotropic distribution of $\bm{B}_{\rm f}$ and introducing the electron spin dephasing time $T_{\Delta}$ due to $\bm{B}_{\rm f}$, we have
\begin {equation}
\langle S_{z}\rangle=S_{0}\left [ 1 - \frac{2}{3}\frac {T_{\rm s}^2}{T_\Delta^2 + T_{\rm s}^2}\mathcal{L} \left(\omega_{\rm e}; \sqrt{\frac{1}{T_\Delta^2} + \frac{1}{T_{\rm s}^2}}\right ) \right ]
\label{StadE},
\end {equation}
where $S_{0}$ is the upper limit of electron spin polarization determined by the competition between the efficiencies of spin-selective injection and spin relaxation, $T_{\rm s}$ is the electron spin lifetime determined by the annihilation time and the spin depolarization times~\cite{SupplMat}, $\mathcal{L}(x; w){=}[1 {+} (x/w)^{2}]^{-1}$ is a Lorentzian function of $x$ with width $w$, $\omega_{\rm e}{=}{\rm g}_{\rm e}\mu_{\rm B}(B_z {+} B_{{\rm n},z}^{\rm stat}) / \hbar$ is an electron Larmor frequency, and $\hbar$ and $\mu_{\rm B}$ are the Dirac constant and the Bohr magneton, respectively. The derivation of Eq.~(\ref{StadE}) is given in Ref.~\cite{SupplMat}. 
The dip of $\langle S_z \rangle$ at $\omega_{\rm e}{\sim}0$, characterized by $T_{\rm s}$ and $T_\Delta$, indicates a more than one order of magnitude narrower $\omega_{\rm e}$-dependence than that of 
the NSP formation rate $\propto \mathcal{L}(\omega_{\rm e}; \tau_{\rm c}^{-1})$ specified by the hyperfine correlation time $\tau_{\rm c}$.
This is because, although $\tau_{\rm c}$ is expected to be tens to hundreds of picoseconds~\cite{Urbaszek13}, both $T_\Delta$~\cite{Kaji19, Braun05, Krebs07, Pal07, Kaji12, Bechtold15} and $T_{\rm s}$~\cite{Watanuki05, Kumano06} are evaluated to be of the order of nanoseconds.
Accordingly, the reduction of the NSP formation rate due to the $\langle S_z \rangle$ deterioration appears as a dip around the compensation condition.
This makes it 
difficult to overleap the compensation condition in the dip region.
Therefore, we expect saturation in the degree of NSP, which is responsible for the Mid.-branch.

With the aforementioned consideration, we calculated $P_{\rm exc}$ dependences of $\langle S_z \rangle$ and $\langle I_z \rangle$, as shown in Fig.~\ref{Fig_Calc}(a).
Here, the magnitude of $P_{\rm exc}$ is expressed by $f_{\rm e}$, an occupation factor of QD by an unpaired electron spin, instead.
While the conventional model (the gray curve in the figure), which does not include the effects of $\bm{B}_{\rm f}$, reproduces a bistable behavior only, the proposed model successfully reproduces the Mid.-branch accompanying a large reduction of $|\langle S_z \rangle |$.
Furthermore, the relation $|B_{{\rm n},z}^{\rm stat}|{<}|B_z|$ in the Mid.-branch is confirmed because the corresponding curve lies above the dotted horizontal line, which indicates the compensation condition $B_{{\rm n},z}^{\rm stat} / (-B_z){=}1$.
These features were consistent with the experimental observations shown in Fig.~\ref{Fig_Experiment1}(b). 
It should be noted that the smaller $|\langle S_z \rangle|$ in the Hi.-branch compared with that in the Lo.-branch cannot be reproduced only by the effect of $\bm{B}_{\rm f}$.
Therefore, we introduced the HFI-induced electron spin-flip relaxation into the calculations~\cite{SupplMat}.
The incorporation of both effects into the model yields better agreements between the experimental [Fig.~\ref{Fig_Experiment1}(b)] and the theoretical [Fig.~\ref{Fig_Calc}(a)] results.

Next, we discuss the transformation from the bistable to the tristable cases with our model.
Figure~\ref{Fig_Calc}(b) shows $f_{\rm e}$ dependence of $\langle I_z \rangle$ with various $\tau_{\rm c}$.
As $\tau_{\rm c}$ increases, the presence of the Mid.-branch becomes less obvious even though the magnitude of $\bm{B}_{\rm f}$ is maintained at the same value in the series of calculations.
Consequently, the curve shape approaches the well-known bistable curve as $\tau_{\rm c}$ is prolonged.
Namely, in the long-$\tau_{\rm c}$ limit, the Mid.-branch is buried in the large hysteresis loop of bistability and experimental access may become difficult.
This trend is consistent with the following observation: QD3 with a large hysteresis loop [Fig.~\ref{Fig_Experiment2}(b)] indicated a bistable behavior [Fig.~\ref{Fig_Experiment2}(a)], while QD2 with a small hysteresis [Fig.~\ref{Fig_Experiment2}(c)] showed the Mid.-branch [Fig.~\ref{Fig_Experiment1}(b)].

Finally, we discuss the possibility of tristability where three stable branches exist \textit{simultaneously}.
As shown in Fig.~\ref{Fig_Calc}(b), the $f_{\rm e}$ range where the Lo., Mid., and Hi.-branches emerge depends strongly on $\tau_{\rm c}$.
Figure~\ref{Fig_Calc}(c) presents the calculated $\langle I_z \rangle$ with $\tau_{\rm c}{=}21$ ps.
In the highlighted region, all stable branches (Lo., Mid., and Hi.) coexist.
Although such a tristability has not been found experimentally thus far, it can be realized if the QD system satisfies specific condition.

In conclusion, we found a new stable branch of NSP in single InAlAs SAQDs.
That implies that NSP indicates three stable branches, although the number of the branches has been believed to be two at the maximum thus far.
The phenomenon was tested by changing the excitation power, as well as polarization, under a longitudinal magnetic field.
The phenomenological model based on the dynamic formation of NSP was developed, including the effect of nuclear spin fluctuation, which successfully explained the three stable branches observed. These observed stable branches are considered to be a general property of NSP in various QD systems. 
Furthermore, the model predicts the tristability of NSP, which may lead to new strategies to prepare complicated QD systems, such as those involving chaotic behavior by using high degrees of freedom.

\begin{acknowledgments}
 This work is supported by JSPS KAKENHI (Grants No. JP26800162 and No.JP17K19046) 
\end{acknowledgments}


\end{document}